\pdfoutput=1
\documentclass[aps,prd,twocolumn,groupaddress,tightenlines,nofootinbib]{revtex4}
\usepackage{graphicx}
\usepackage{epsfig}
\usepackage{bm}
\usepackage{latexsym,amssymb,amsmath,amsfonts,amssymb,txfonts,pxfonts,wasysym,float}
\usepackage{color}

\newcommand{\beq}[1]{\begin{equation}\label{#1}}
\newcommand{\eeq}{\end{equation}}
\newcommand{\bea}[1]{\begin{eqnarray} \label{#1}}
\newcommand{\eea}{\end{eqnarray}}
\newcommand{\ba}{\begin{array}}
\newcommand{\ea}{\end{array}}

\def\be{\begin{equation}}
\def\ee{\end{equation}}
\def\gs{\mathrel{
   \rlap{\raise 0.511ex \hbox{$>$}}{\lower 0.511ex \hbox{$\sim$}}}}
\def\ls{\mathrel{
   \rlap{\raise 0.511ex \hbox{$<$}}{\lower 0.511ex \hbox{$\sim$}}}}

\newcommand{\postscript}[2]{\setlength{\epsfxsize}{#2\hsize}
   \centerline{\epsfbox{#1}}}

\newcommand{\comment}[1]{}

\usepackage[usenames,dvipsnames]{xcolor}
\definecolor{orange}{cmyk}{0,0.5,1,0}
\definecolor{rossoCP3}{cmyk}{0,.88,.77,.40}
\definecolor{graa}{rgb}{0.8,0.8,0.8}
\definecolor{blaa}{rgb}{0.2,0.2,0.6}

\begin{document}

\title{\color{rossoCP3}{Acceleration of ultrahigh-energy cosmic rays in starburst superwinds}}

\author{Luis Alfredo Anchordoqui}
\affiliation{Department of Physics \& Astronomy,  Lehman College, City University of
  New York, NY 10468, USA \\
Department of Physics,
 Graduate Center, City University
  of New York,  NY 10016, USA\\
Department of Astrophysics,
 American Museum of Natural History, NY
 10024, USA\\
Departamento de F\'{\i}sica,
Universidad Nacional de La Plata,  (1900) La Plata, Argentina
}

\begin{abstract}
  \noindent The sources of ultrahigh-energy cosmic rays (UHECRs) have
  been stubbornly elusive. However, the latest report of the Pierre
  Auger Observatory provides a compelling indication for a possible
  correlation between the arrival directions of UHECRs and nearby
  starburst galaxies. We argue that if starbursts are sources of
  UHECRs, then particle acceleration in the large-scale terminal shock
  of the superwind that flows from the starburst engine represents the
  best known concept model in the market. We investigate new
  constraints on the model and readjust free parameters
  accordingly. We show that UHECR acceleration above about
  $10^{11}~{\rm GeV}$ remains consistent with observation. We also
  show that the model could accommodate hard source spectra as
  required by Auger data. We demonstrate how neutrino emission can be
  used as a discriminator among acceleration models.
\end{abstract}

\maketitle

The search for the sources of ultrahigh-energy cosmic rays (UHECRs)
remains one of the cornerstone components of high energy astrophysics.
The source hunting exploration is mostly driven by three observables:
the energy spectrum, the nuclear composition, and the distribution of
arrival directions. From these observables, the last one allows the
most direct conclusions about the locations of UHECR accelerators.

Very recently, the Pierre Auger Collaboration reported an indication
of a possible correlation between UHECRs ($E > 10^{10.6}~{\rm GeV}$)
and nearby starburst galaxies, with an {\it a posteriori} chance
probability in an isotropic cosmic ray sky of $4.2 \times 10^{-5}$,
corresponding to a 1-sided Gaussian significance of
$4\sigma$~\cite{Aab:2017njo}. The smearing angle and the anisotropic
fraction corresponding to the best-fit parameters are $13^\circ$ and
10\%, respectively.  The energy threshold coincides with the observed
suppression in the
spectrum~\cite{Abbasi:2007sv,Abraham:2008ru,Abraham:2010mj,AbuZayyad:2012ru}. Interestingly,
when we properly account for the barriers to UHECR propagation in the form
of energy loss
mechanisms~\cite{Greisen:1966jv,Zatsepin:1966jv,Puget:1976nz} we
obtain a self consistent picture for the observed UHECR horizon.

On a separate track, the Telescope Array Collaboration has reported an
intriguing excess of UHECRs ($E > 10^{10.76}~{\rm GeV}$) above the
isotropic background-only expectation, with a chance probability of
$3.7 \times 10^{-4}$, corresponding to
$3.4\sigma$~\cite{Abbasi:2014lda,Kawata:2015whq}.  This {\it hot spot}
spans a $\sim 20^\circ$ region of the sky, and the starburst galaxy
M82 is close to the best-fit source position~\cite{He:2014mqa,Anchordoqui:2002dj}.

In this paper we argue that if starbursts are sources of UHECRs, then
particle acceleration in the large-scale terminal shock of the
superwind that emanates from the starburst
nucleus~\cite{Anchordoqui:1999cu} represents the best known concept
model in the market. We investigate new constraints on the model and
readjust free parameters accordingly. We show that acceleration of
UHECR nuclei in the range $10^{10.6} \alt E/{\rm GeV} \alt 10^{11}$
remains consistent with the most recent astrophysical observations.
We also show that the model could accommodate hard source spectra as
required by Auger data.

Extremely fast spinning young pulsars~\cite{Blasi:2000xm,Fang:2012rx},
newly born magnetars~\cite{Arons:2002yj}, gamma-ray bursts
(GRBs)~\cite{Waxman:1995vg,Vietri:1995hs}, and tidal disruption events
(TDEs) caused by  black holes~\cite{Farrar:2008ex} have
been identified as potential UHECR accelerators inside starburst galaxies~\cite{Aab:2017njo,Pfeffer:2015idq}. Given the ubiquity of
pulsars, magnetars, and  black holes we can ask ourselves
why the correlation of UHECRs with starburst galaxies would be
explained by the presence of these common objects. Rather there must
be some other inherently unique feature(s) of starburst galaxies to
account for this correlation. A true smoking gun for the
pulsar/magnetar/TDE scenario would be a correlation with the
distribution of {\it all} nearby matter as opposed to a particular
class of objects~\cite{note1}.

There are numerous indications that long GRBs are extreme
supernova events, which arise from the death of massive
stars~\cite{MacFadyen:1998vz}.  Starburst galaxies are characterized
by high star-formation rates per unit area, of the order of 15 to
20~$M_\odot~{\rm yr^{-1}} \, {\rm kpc}^{-2}$~\cite{Heckman}. This is up to several hundred times larger
than the characteristic value normally found in gas-rich 
galaxies like the Milky Way.  The observed supernova rate in
starbursts is also higher than average, and so it seems only natural to
expect a high rate of long GRBs too~\cite{Dreyer:2009pj,Biermann:2016xzl}. However, the
star formation rates per unit stellar mass of GRB host galaxies are
found to be higher than for typical nearby starburst
galaxies~\cite{Chary:2001yx}. Moreover, stronger and stronger
experimental evidence has been accumulating that implies GRB hosts are low mass
irregular galaxies and have low metallicity, see
e.g.~\cite{Stanek:2006gc,Modjaz:2007st,Jimenez:2013dka}. Altogether,
this makes the GBR $\leftrightharpoons$ (metal-rich) starburst
connection highly unlikely.

The universal fast star formation in starburst galaxies is directly
correlated with the efficient ejection of gas, which is the fuel for
star formation. This happenstance generates a galactic-scale
superwind, which is powered by the momentum and energy injected by
massive stars in the form of supernovae, stellar winds, and
radiation~\cite{Heckman,Veilleux:2005ia}. Multi-wavelength
observations seem to indicate that these superwinds are genuinely
multi-phase: with hot, warm, cold, and relativistic (cosmic rays)
phases. These observations also suggest a pervasive development of the
hot ($T \sim 10^{7}~{\rm K}$) and warm diffuse ionized ($T \sim
10^4~{\rm K}$) phases. Namely, experiment shows that the hot and warm
large-scale supersonic outflows escalate along the rotation axis of
the disk to the outer halo area in the form of local chimneys.  Such a
supersonic outflow, however, does not extend indefinitely. As the
superwind expands adiabatically out beyond the confines of the
starburst region, its density decreases. At a certain radial distance
the pressure would become too small to further support a supersonic
flow. Whenever the flow is slowed down to subsonic speed a termination
shock stops the superwind. The shocked gas continues as a subsonic
flow. The termination shock would remain in steady state as long as
the starburst lasts. As noted elsewhere~\cite{Anchordoqui:1999cu} this
set up provides a profitable arena for acceleration of UHECRs.

Next, in light with our stated plan, we examine new constraints on the 
model. Consider a spherical cavity where core-collapse supernovae and
stellar winds inject kinetic energy. This kinetic energy then
thermalizes and drives a super-heated outflow that escapes the
sphere. Following~\cite{Chevalier:1985pc}, to a first approximation we
ignore gravity, radiative cooling, and other effects. In this
approximation energy conservation leads to the asymptotic speed of the
outflow
\begin{equation}
v_\infty \approx \sqrt{\frac{2 \dot  E_{\rm sw}}{\dot M_{\rm sw}}} \sim 10^3
\sqrt{\frac{\epsilon}{\beta}}~~{\rm km} \ {\rm s^{-1}} \,, 
\label{vinfinity}
\end{equation}
where $\dot E_{\rm sw}$ and $\dot M_{\rm sw}$ are respectively the energy and mass
injection rates inside the spherical volume of the
starburst region, and where
$\beta$ is the mass loading factor, i.e. the ratio of the mass
injection rate to the star formation rate.  In the
second rendition we have scaled the energy injection rate expected
from core-collapse supernovae considering a thermalization efficiency
$\epsilon$.  For this order of magnitude calculation, we have assumed
that in total a $100 M_\odot$ star injects ${\cal O} (10^{51}~{\rm
  erg})$ into its surroundings during the wind phase. 

As the cavity expands adiabatically a strong shock front is formed on
the contact surface with the cold gas in the halo. At the region where
this occurs, the inward ram pressure is balanced by the pressure
inside the halo, $P_{\rm halo}$. A point worth noting at this juncture is that the
difference in pressure between the disk and the halo manifestly breaks the
symmetry, and so the outflowing fluid which escapes from the starburst
region features back-to-back chimneys with conic
profiles. Rather than considering a spherical shock we assume the outflow
cones fill a solid angle $\Omega$, and hence the ram pressure
at radius $r$ is found to be
\begin{equation}
P_{\rm ram} = \frac{\rho_{\rm sw} \  v_\infty^2}{2} = \frac{\dot
  p_{\rm sw}}{2 \Omega \ r^2} = \frac{\dot M_{\rm sw} \
  v_\infty}{2 \Omega \ r^2} = \frac{\sqrt{2 \ \dot E_{\rm sw} \
    \dot M_{\rm sw} }
}{2 \Omega \ r^2}  \,,
\end{equation}
where $\rho_{\rm sw} = \dot M_{\rm sw}/(\Omega v_\infty r^2)$ is the density of the
outflow and $\dot p_{\rm sw} = \dot M_{\rm sw} \, v_\infty$ is the
asymptotic momentum injection rate of the
superwind~\cite{Lacki:2013zsa}. The  agitated superwind gas inside the
shock is in pressure equilibrium with the outside gas at a  radius
\begin{equation}
R_{\rm sh} \sim \sqrt{\frac{\dot M_{\rm sw} \ v_\infty }{2 \Omega \ P_{\rm halo}}}
\, . 
\end{equation}
The termination shock is a steady-state feature, present even if the
starburst wind has always been active.

\begin{figure}[tbp] 
    \postscript{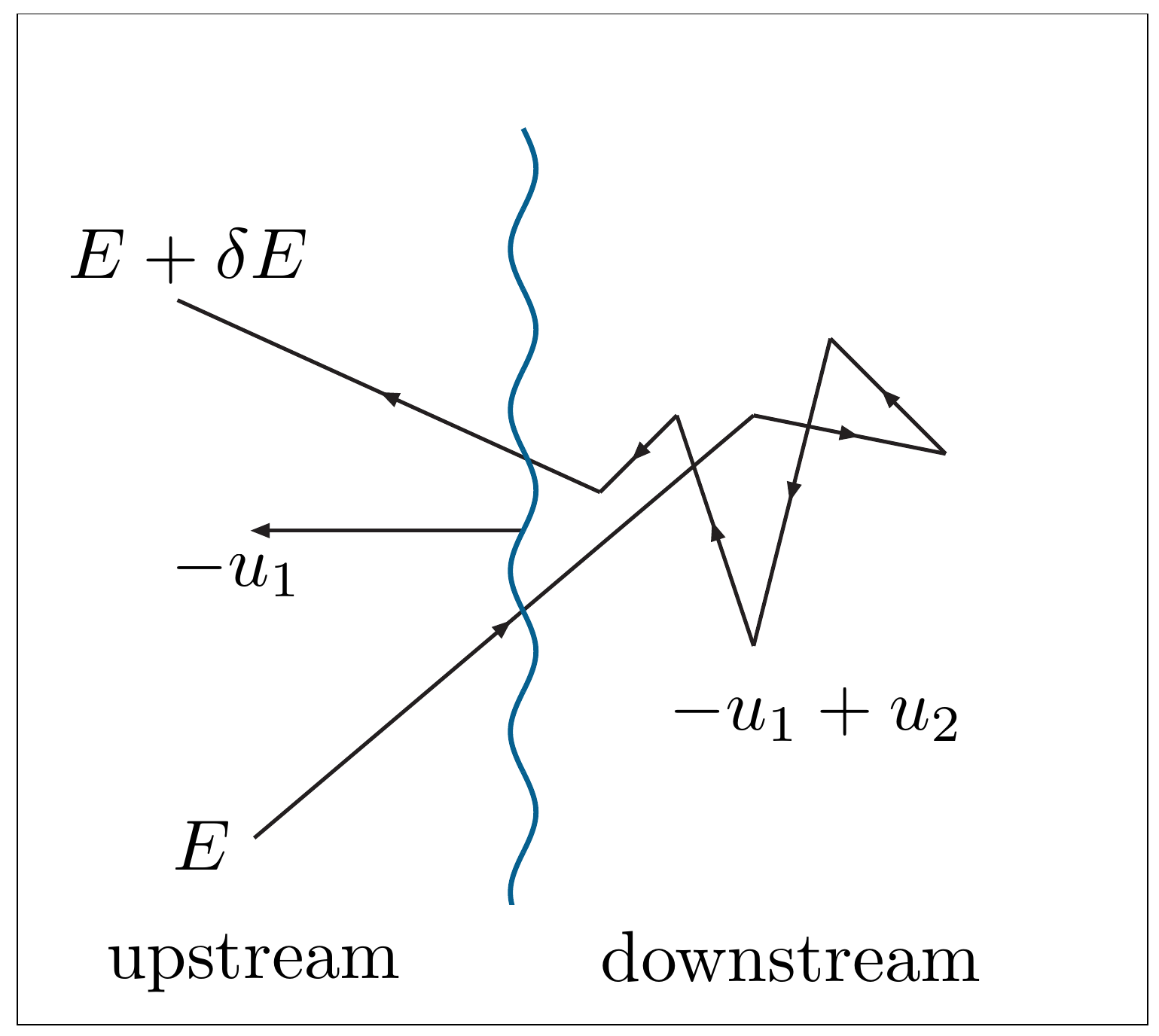}{0.9} 
    \caption{An sketch of diffusive shock acceleration.  A plane shock
      front moves with velocity $- u_1$. The shocked gas flows away 
      from the shock with a velocity $u_2$ relative to the shock
      front, where $|u_2| < |u_1|$. This implies that in the lab frame the gas
      behind the shock moves to the left with velocity $-u_1 +
      u_2$. It is easily seen that the average energy gain per
      encounter $\xi = \langle \delta E \rangle/E = 4 (u_1 -
      u_2)/3$. }
\label{fig} 
\end{figure}

All told, we expect relativistic baryons of charge $Ze$, which could
be dragged from the starburst core into the superwind, to experience
diffusive shock
acceleration~\cite{Bell:1978zc,Bell:1978fj,Lagage:1983zz,Drury:1983zz,Blandford:1987pw}.
Diffusive shock acceleration is a first-order Fermi acceleration
process~\cite{Fermi:1949ee} in which charged particles increase their
energy by crossing the shock front multiple times, scattering off
turbulence in the magnetic field $B$, as shown in Fig.~\ref{fig}.  The
magnetic field turbulence is assumed to lead to isotropization and
consequent diffusion of energetic particles which then propagate
according to the standard transport theory. The acceleration time
scale is given by
\begin{equation}
\left(\frac{1}{E} \frac{dE}{dt} \right)^{-1} = \frac{T_{\rm
      cycle}}{\xi} \,,
\end{equation}
where 
\begin{equation}
T_{\rm cycle} = 4 \kappa  \left( \frac{1}{u_1} + \frac{1}{u_2} \right)
\end{equation}
is the cycle time for one back-and-forth encounter,
\begin{equation}
\xi \sim \frac{4}{3} \ (u_1 - u_2) 
\end{equation}
is the fractional energy gain per encounter,
\begin{equation}
\kappa = \frac{1}{3} R_L \sim \frac{1}{3} \frac{E}{ZeB} 
\end{equation}
is the Bohm diffusion coefficient, $u_1 \sim v_\infty$ is the
upstream flow (unshocked gas) velocity, and
$u_2$ the downstream (shocked gas) velocity~\cite{Gaisser:1990vg}. Now, using the continuity of mass flow across the
shock together with the kinetic theory of gases we arrive at the
shock compression ratio
\begin{equation}
\zeta = \frac{u_1}{u_2} \approx  \frac{\gamma +1}{\gamma -1} \,,
\end{equation}
where $\gamma$
is the adiabatic index and where we have assumed the strong shock condition
in which the Mach number of the flow  $\gg 1$~\cite{Lagage:1983zz}.

There exists {\it lore} that convinces us that diffusive shock
acceleration of UHECRs is associated to the adiabatic index of a
monoatomic classic gas $\gamma = 5/3$~\cite{Gaisser:1990vg}. This assumption leads to $\zeta
= 4$. In what follows, we move away from the stereotype and take
$\gamma = 9/7$, which is associated to a three-atomic gas with
non-static bindings. Our assumption gives $\zeta = 8$ and $\xi \sim v_\infty$. The rationale
for this particular choice will be given below. Assuming that the
acceleration is continuous, the constraint due to the finite lifetime
$\tau$ of the shock yields,
\begin{equation}
E_{\rm max} \sim \frac{1}{12} \ Ze \  B \ v_\infty^2 \
\tau \, .
\label{Emax}
\end{equation} 
Before proceeding, we note that the rate of acceleration for our
choice of $\gamma =9/7$ is slower by a factor of 1.8 when compared to the
rate for $\gamma = 5/3$, and consequently $E_{\rm max}$ is reduced.
In the preceding discussion it was implicitly assumed that the magnetic
field is parallel to the shock normal.  Injecting additional
constraints into the model may reduce the maximum achievable
energy~\cite{Bustard:2016swa}.

To develop some sense of the orders of magnitude involved, we assume
that the prominent M82 typifies the nearby starburst population. For a
standard Kroupa initial mass function~\cite{Kroupa:2002ky}, our
archetypal starburst has a star formation rate $\sim 10 M_\odot~{\rm
  yr}^{-1}$ and a radius of about $400~{\rm pc}$. Hard X-ray
observations provide direct observational evidence for a hot-fluid
phase.  The inferred gas temperature range is $10^{7.5} \alt T/{\rm K}
\alt 10^{7.9}$, the thermalization efficiency $0.3 \alt \epsilon \alt
1$, and the mass loading factor $0.2 \alt \beta \alt
0.6$. Substituting for $\epsilon$ and $\beta$ into (\ref{vinfinity})
we obtain $1.4 \times 10^3 \alt v_\infty/({\rm km \, s}^{-1}) \alt 2.2
\times 10^3$~\cite{Strickland:2009we}. The warm fluid has been
observed through nebular line and continuum emission in the vacuum
ultraviolet, as well as through mid- and far-infrared fine-structure
line emission
excitations~\cite{Heckman:2000sj,Hoopes:2006mz,Beirao,Contursi:2012wa}. High-resolution
spectroscopic studies seem to indicate that the warm ($T \sim
10^4~{\rm K}$) gas has emission-line ratios consistent with a mixture
of photo-ionized gas by radiation leaking out of the starburst and
shock-heated by the outflowing superwind fluid generated within the
starburst~\cite{Heckman:1990fe}.  The kinematics of this gas, after
correcting for line-of-sight effects, yields an outflow speed of the
warm ionized fluid of roughly 600~${\rm km} \, {\rm s}^{-1}$. The
velocity field, however, shows rapid acceleration of the gas from the
starburst itself out to a radius of about 600~pc, beyond which the
flow speed is roughly constant.  The inferred speed from cold and warm
molecular and atomic gas observations~\cite{Veilleux:2009rb,Leroy} is
significantly smaller than those observed from the warm ionized phase.
This is also the case for the starburst galaxy NGC 253: ALMA
observations of CO emission imply a mass loading factor of at least 1
to 3~\cite{Bolatto:2013aqa}. However, it is important to stress that
the emission from the molecular and atomic gas most likely traces the
interaction of the superwind with detached relatively denser ambient
gas clouds~\cite{Heckman}, and as such it is not the best gauge to
characterize the overall properties of the superwind
plasma~\cite{Lacki:2013sda}. (See~\cite{Gustavo} for a different
perspective.) Herein, we adopt the properties of the hot gas detected
in hard X-rays to determine the shock terminal velocity. We take an
outflow rate of $\dot M_{\rm sw} \sim 3 M_\odot~{\rm yr}^{-1}$, which
is roughly 30\% of the star-formation rate ($\beta \sim 0.3$),
yielding $\dot E_{\rm sw} \sim 3 \times 10^{42}~{\rm erg}\,
s^{-1}$~\cite{Heckman}. For $\Omega \sim \pi$, this leads to $v_\infty
\sim 1.8 \times 10^3~{\rm km \ s}^{-1}$ and $R_{\rm sh} \sim 8~{\rm
  kpc}$, where we have taken $P_{\rm halo} \sim 10^{-14}~{\rm erg} \,
{\rm cm}^{-3}$~\cite{Shukurov:2002hh}.

Radio continuum and polarization observations of M82 provide an
estimate of the magnetic field strength in the core region of
$98~\mu{\rm G}$ and in the halo of $24~\mu{\rm G}$; averaging the
magnetic field strength over the whole galaxy results in a mean
equipartition field strength of $35~\mu{\rm
  G}$~\cite{Adebahr:2012ce}. Comparable field strengths have been estimated for
NGC 253~\cite{Beck,Heesen:2008cs,Heesen:2009sg,Heesen:2011kj} and
other starbursts~\cite{Krause:2014iza}. Actually, the field strengths
could be higher if the cosmic rays are not in equipartition with the
magnetic
field~\cite{Thompson:2006is,Paglione:2012ma,Lacki:2013ry}.  
If this were the case, e.g., the magnetic field strength in M82 and NGC 253
could be as high as $300~\mu$G~\cite{DomingoSantamaria:2005qk,delPozo:2009mh,Lacki:2013nda}.

The duration of the starburst phenomenon is subject to large
uncertainties. The most commonly cited timescale for a starburst is 5
to 10~Myr, comparable to the lifetime of massive
stars~\cite{Thornley:2000ib,Torres:2004ui,Torres:2012xk}. However, it has been suggested that the
starburst phenomenon can be a longer and more global event than
related by the lifetime of individual massive stars or pockets of
intense star formation~\cite{Meurer:2000yq,McQuinn:2009gc,McQuinn:2010kn}. In this
alternative viewpoint the short duration timescales are instead
interpreted as a measure of the {\it flickering} created by currently
active pockets of star formation that move around the
galaxy. Measuring the characteristics of just one of these flickers
reveals much about an individual star formation region but of course
does not measure the totality of the starburst phenomenon in the
galaxy. If starbursts are indeed a global phenomenon, then the events
are longer than the lifecycle of any currently observable massive star
or area of intense star formation and the bursts are not
instantaneous. An observation that measures currently observable star
formation activity will therefore only measure the {\it flickering}
associated with a starburst pocket and not the entire phenomenon. This
aspect, frequently denied or not yet sufficiently emphasized, may
bring still another rewarding dimension to the problem at hand.

A measurement of the starburst phenomenon in twenty nearby galaxies
from direct evaluation of their star formation histories reconstructed
using archival Hubble Space Telescope observations suggests the average
duration of a starburst  is between 450 and 
650~Myr~\cite{McQuinn:2010kn}. 

Since the large-scale terminal shock is far from the starburst region,
the photon field energy density in the acceleration region drops to
values of the order of the cosmic microwave background. Now,
for $E \alt 10^{11}~{\rm GeV}$ and $Z \agt 10$, the energy attenuation
length $\agt 30~{\rm Mpc}$~\cite{Allard:2011aa}. Therefore, we will
restrict ourselves to $\tau \alt 100~{\rm Myr}$. This duration range
is in good agreement with the overall star formation history of
M82~\cite{deGrijs:2001ec,deGrijs:2002qk} and NGC
253~\cite{Davidge:2010qg,Davidge}, and it is also consistent with the
upper limit on the starburst age of these galaxies derived
in~\cite{Rieke:1980xt}.

In toto, substituting $v_\infty \sim 1.8 \times 10^3~{\rm km \, s}^{-1}$, $B \sim 300~\mu$G, and $\tau  \sim 40~{\rm Myr}$
into (\ref{Emax}) we obtain
\begin{equation}
E_{\rm max}
\sim Z \,  10^{10}~{\rm GeV} \, .
\label{Emax2}
\end{equation}
Note  that
(\ref{Emax2}) is consistent with the Hillas
criterion~\cite{Hillas:1985is}, as the maximum energy of confined
baryons at a shock distance of $R_{\rm sh}$ is found to be
\begin{equation}
E_{\rm max} \simeq 10^9 \ Z \ \frac{B}{\mu G} \ \frac{R_{\rm sh}}{{\rm
    kpc}}~{\rm GeV} \ .
\end{equation}
The shape of the source emission spectrum  is then driven by
UHECR leakage from the boundaries of the shock (a.k.a. direct
escape). 

Next, we generalize the scaling arguments for direct escape given
in~\cite{Baerwald:2013pu} to provide a justification for our choice of
the adiabatic index of a polyatomic gas.  Consider an expanding shell
that magnetically confines UHECR nuclei. Assuming that the nuclei are
isotropically distributed in the shell, the number of escaping
particles is proportional to the volume. The shell width expands as
$\delta r \propto r$. This implies that the volume of the
plasma increases as $V \propto r^3$ and the total energy scales as $U \propto
V^{-(\gamma -1)} \propto r^{-3 (\gamma -1)}$. Now, using the scaling
of the volume and the total energy we derive the scaling of the
magnetic field inside the plasma $B \propto \sqrt{U/V} \propto r^{-3
  \gamma/2}$. If we further assume that the energy of a single
particle in the plasma scales in the same way as the total energy of
the plasma, then the Larmor radius of the particle changes with time
(or radius) as $R_L \propto E/B \propto r^{-3 (\gamma -2)/2}$. For a
relativistic gas, $\gamma = 4/3$ yielding $R_L \propto r$, and so 
the ratio $R_L/\delta r$  is constant.  This means that a relativistic gas
provides a critical balance for stability between
losses and escape. For $\gamma > 4/3$, the adiabatic energy loss is
faster than the escape, and the particles are more strongly confined
for larger radii.  For $\gamma < 4/3$, the Larmor radius increases faster
than the particles lose energy, and the particles are getting less
confined at larger radii. Now, the main prediction of diffusive shock
acceleration is that the final cosmic ray distribution function is a
power-law function in momentum space $f(p) \propto p^{-3\zeta /(\zeta
  - 1)}$~\cite{Lagage:1983zz}.  The source energy spectrum $N(E)
\propto E^{-\alpha}$ is related to the momentum spectrum by $N(E) \ dE
= f(p) \ 4 \pi \ p^2 \ dp$. Interestingly, for $\gamma = 9/7$ we obtain a
hard source spectrum, with spectral index $\alpha = 1.4$. Note that
simultaneously reproducing Auger data on the spectrum together with
the observed nuclear composition also requires hard 
spectra at the sources~\cite{Aloisio:2013hya,Unger:2015laa,Aab:2016zth}.

At this stage, we pause to compare our results with those
in~\cite{Bustard:2016swa,Gustavo}: {\it (i)}~The efficiency of the
acceleration process (i.e., the normalization constant of the
acceleration rate) is reduced by factor of 1.8 in our calculations due
to a larger shock compression ratio. {\it (ii)}~The fiducial value of
the magnetic field strength in the halo adopted in this paper is a
factor of 375 larger than the one considered in~\cite{Bustard:2016swa}
and a factor of 60 larger than the one in~\cite{Gustavo}.  {\it
  (iii)}~The duration of the starburst phenomenon considered in this work is a
factor of 2.5 smaller than the one adopted in~\cite{Bustard:2016swa}
and a factor of 4 larger than that adopted in~\cite{Gustavo}. The
magnetic field strength considered herein is supported by
multi-frequency observations~\cite{DomingoSantamaria:2005qk,delPozo:2009mh,Lacki:2013nda}. The duration of
the starburst phase is based on the hypothesis that the
non-equilibrium energy output and mass transfer from an individual
pocket of star formation may only impact the local star cluster
without shutting down the bursting phenomenon, which to first order is
not self-quenching; this is also supported by
experiment~\cite{McQuinn:2010kn}. Spanning the allowed
range $\tau$ could be up to a factor of 2.5 larger, relaxing the
requirements on $B$ and $v_\infty$.

The Galactic magnetic field is not well constrained by current data, but if we
adopt recent models~\cite{Pshirkov:2011um,Jansson:2012pc,Jansson:2012rt,Unger:2017kfh},
typical values of the deflections of UHECRs crossing the Galaxy are
\begin{equation}
\theta \sim 10^\circ \ Z  \ \left(\frac{E}{10^{10}~{\rm GeV}}
\right)^{-1} \,,
\label{deflection}
\end{equation} 
depending on the direction
considered~\cite{Farrar:2017lhm,Aab:2017tyv}. To account for the
potential anisotropic signal, which spans the energy range $10^{10.6}
\alt E/{\rm GeV} \alt 10^{11}$, (\ref{deflection}) argues in favor of
baryonic UHECR with $Z \alt 10$, in agreement with the nuclear
composition observed in this energy
range~\cite{Aab:2014kda,Aab:2014aea,Aab:2016htd,Abbasi:2015xga}.  

In closing, we note  that if a source produces an
anisotropy signal at energy $E$ with cosmic ray nuclei of charge $Ze$,
it should also produce a similar anisotropy pattern at energies $E/Z$
via the proton component that is emitted along with the nuclei, given
that the trajectory of cosmic rays within a magnetic field is only
rigidity-dependent~\cite{Lemoine:2009pw}. To suppress the accompanying
proton flux we follow~\cite{Anchordoqui:2017abg} and assume that the
relativistic flux of nuclei that is dragged into the starburst
superwind originates in the surface of newly born
pulsars~\cite{Fang:2012rx,note2}. As noted in~\cite{Liu:2013ppa}, secondary
protons produced during propagation could also create an anisotropy
pattern in the ``low'' energy regime. This sets a constraint on the
maximum distance to nucleus-emitting-sources.  Making the extreme
assumption that the source does not emit any proton, the source(s)
responsible for the suggested anisotropies should lie closer than
$\sim 20$ to 30, 80 to 100, and 180 to 200~Mpc, if the anisotropy
signal is mainly composed of oxygen, silicon and iron nuclei,
respectively~\cite{Liu:2013ppa}. This sets an interesting constraint
on the model and provides a distinctive signal to be tested by future
data.

In summary, we have shown that UHECR acceleration ($10^{10.6} \alt
E/{\rm GeV} \alt 10^{11}$) in the superwind of starburst galaxies
remains consistent with observation.  Even though from an astronomical
perspective starbursts are thought to be a short-lived phenomena,
UHECR acceleration requires longer global starburst durations. The
longer durations would imply that starbursts may not extinguish
themselves through energy and mass transfer, but instead may be
self-regulating environments. If these longer duration and more global
starburst events are typical of bursting galaxies, then the starburst
phenomenon could have a larger impact on galactic evolution than
previously thought.  For example, a long-duration starburst would make
the ratio of baryons to dark matter  drop rapidly with decreasing
halo mass, relaxing the discrepancy between theory and
experiment~\cite{McGaugh:2009mt}. Future data from
AugerPrime~\cite{Aab:2016vlz} and POEMMA~\cite{Olinto:2017xbi} may
confirm the cross-correlation between UHECRs and starbursts,
supporting longer duration global bursts and thereupon extending the
scope of multi-messenger astrophysics. 

We have also shown that the starburst superwind hypothesis could
develop hard source spectra as required by Auger data. These hard
spectra would also have profound implications for the multi-messenger
program. Note that since the maximum energy in the acceleration
process is constrained by direct escape of the nuclei, the flux of
photons and neutrinos accompanying the starburst UHECR emission would
be strongly suppressed.  Interestingly though, we can use the
suppressed emission of ultrahigh-energy neutrinos to differentiate
between UHECR acceleration models. This is because for UHECRs crossing
the supernova ejecta surrounding neutron stars, the effective optical
depth to hadronic interactions is larger than unity, and so even in
the most pessimistic case we expect fluxes of neutrinos in the energy
range $10^8 \alt E_\nu/{\rm GeV} \alt
10^9$~\cite{Fang:2013vla}. Indeed, upper limits on the diffuse
neutrino flux from IceCube~\cite{Aartsen:2013uuv,Aartsen:2013dsm} and
the Pierre Auger Observatory~\cite{Aab:2015kma} already constrain
models of UHECR acceleration in the core of starburst
galaxies~\cite{Fang:2015xhg,Fang:2017mhl}. Note, however,
that if high-energy cosmic rays are re-accelerated to ultrahigh
energies at the terminal shock of the starburst superwind, we expect
the neutrino emission from starbursts to cutoff somewhat above
$10^7~{\rm GeV}$, as entertained in~\cite{Loeb:2006tw}.

\acknowledgments{I would like to acknowledge many useful discussions
  with Vernon Barger, Gustavo Romero, Diego Torres, Michael Unger, Tom Weiler, and
  my colleagues of the Pierre Auger and POEMMA collaborations. I would
  also like to thank Carlos Garcia Canal and the Department of Physics
  at the UNLP for hospitality. This work has been
  supported by the U.S. National Science Foundation (NSF) Grant
  No. PHY-1620661 and by the National Aeronautics and Space
  Administration (NASA) Grant No. NNX13AH52G. The views and opinions expressed in
  this article are solely those of the author.}

\end{document}